\documentclass[12pt, journal,onecolumn]{IEEEtran}
\usepackage{times}

\usepackage{amsmath}
\usepackage{amsfonts}
\usepackage{latexsym}
\usepackage{amssymb}
\usepackage{amsthm}
\usepackage{upref}
\usepackage{graphicx}
\usepackage{psfrag}

\usepackage{enumerate} 
\usepackage{xcolor}
\newtheorem{thm}{Theorem$\!$}

\newtheorem{lem}[thm]{Lemma$\!$}

\newtheorem{prop}[thm]{Proposition$\!$}

\newtheorem{cor}[thm]{Corollary$\!$}

\newtheorem{defn}[thm]{Definition$\!$}

\newtheorem{xmpl}[thm]{Example$\!$}

\newtheorem{cnstr}[thm]{Construction$\!$}

\newtheorem{algr}[thm]{Algorithm$\!$}

\newtheorem{conj}[thm]{Conjecture$\!$}

\usepackage{setspace}
\usepackage[margin=1in]{geometry}
\begin{document}
\doublespacing
\title{Machine Learning for Error Correction \\ with Natural Redundancy}


\author{Pulakesh Upadhyaya,~\IEEEmembership{Student Member,~IEEE} \\ 
Anxiao (Andrew) Jiang,~\IEEEmembership{Senior Member,~IEEE}
\thanks{
 This work was supported in part by NSF Grant CCF-1718886. Parts of this paper were presented at the 2017 Allerton Conference on Communication, Control, and Computing, 2017 Information Theory and Applications (ITA) Workshop, and 2019 IEEE International Conference on Communications (ICC). The authors are with the Department of Computer Science and Engineering, Texas A\&M University, College Station, TX 77840, USA (e-mail: pulakesh@tamu.edu; ajiang@cse.tamu.edu). This paper was submitted to IEEE Journal on Selected Areas in Information Theory (special issue on Deep Learning: Mathematical Foundations and Applications to Information Science).}}

\maketitle

\begin{abstract}
The persistent storage of big data requires advanced error correction schemes. The classical approach is to use error correcting codes (ECCs). This work studies an alternative approach, which uses the redundancy inherent in data itself for error correction. This type of redundancy, called Natural Redundancy (NR), is abundant in many types of uncompressed or even compressed files. The complex structures of Natural Redundancy, however, require machine learning techniques. In this paper, we study two fundamental approaches to use Natural Redundancy for error correction. The first approach, called Representation-Oblivious, requires no prior knowledge on how data are represented or compressed in files. It uses deep learning to detect file types accurately, and then mine Natural Redundancy for soft decoding. The second approach, called Representation-Aware, assumes that such knowledge is known and uses it for error correction. Furthermore, both approaches combine the decoding based on NR and ECCs. Both experimental results and analysis show that such an integrated scheme can substantially improve the error correction performance.

%This paper studies how to use machine learning techniques to utilize natural redundancy in data for error correction, and how to combine it with error-correcting codes to effectively improve data reliability. Two different schemes are discussed. The first  scheme is representation-oblivious: it requires no prior knowledge on how data are represented (e.g., mapped from symbols to bits, compressed, and combined with meta data) in different types of files, which makes the solution more convenient to use for storage systems. The second scheme is representation aware, where the error correction NR decoder has prior knowledge of the compression algorithms. In both cases, the natural redundancy decoder is combined with an ECC decoder to improve the error correction performance. 
\end{abstract}

\IEEEpeerreviewmaketitle

\section{Introduction}
A large amount of data is generated on the Internet everyday, and the feasibility of storing useful data permanently has become a key concern. The  most effective classic approach to improve data reliability is to add external redundancy to data using Error Correcting Codes (ECCs). We call such redundancy \emph{artificial redundancy}. However, over time, errors accumulate in storage systems and can exceed the decoding threshold of ECCs. To ensure permanent reliability of data, many techniques have been explored to improve the error correction capabilities in long-term storage systems. Recent progress in machine learning has offered an opportunity to employ novel techniques to improve data reliability. One such approach is to use Natural Redundancy in data for error correction. 

By \emph{Natural Redundancy} (NR), we refer to the redundancy that is inherent in
data, which is not artificially added by ECCs. It is abundant in many types of uncompressed or even compressed files. For instance, consider the English language.  When LZW (Lempel-Ziv-Welch) coding is used with a fixed dictionary of $2^{20}$ patterns (larger than many LZW codes in practical systems), the language can be compressed to 2.94 bits/character. State-of-the art compression algorithms (e.g., syllable-based Burrows-Wheeler Transform) can further reduce it to 2 bits/character~\cite{Lansky2007}. However, even with such advanced compression techniques, the result is still far from Shannon's estimation of 1.34 bits/character, which is an upper-bound for the entropy of printed English~\cite{Shannon1951English}. For images, residual redundancy can also be abundant after compression, as made evident by recent inpainting techniques of deep learning~\cite{Xie}. Such abundant Natural Redundancy can be an excellent resource for error correction. 

There are two fundamental ways to utilize Natural Redundancy in an information system. The first way is enhanced \textit{data compression}, which often uses deep learning to remove redundancy further than before~\cite{liDatacompression,prakashDatacompression}. It is a new and active research area, and compression ratios higher than classic compression algorithms have been achieved in some cases (e.g., for high distortion regimes).

The second way, which is the focus of this paper, is to use Natural Redundancy for \textit{error correction}. That is, a new decoder is designed to mine the Natural Redundancy in data, and utilize it for error correction. The decoder can be further combined with ECC's decoder for better performance. A strong motivation for this method is that modern storage systems already store a massive amount of data, which would be very costly to reprocess. The Natural Redundancy (NR) based decoder does not require systems to examine or modify any existing data. It only requires an enhancement to the decoding algorithm itself. Therefore, it is compatible with storage systems and convenient to use. 

In this paper, we study two fundamental approaches to use Natural Redundancy for error correction. The first approach, called \textit{Representation-Oblivious}, requires no prior knowledge on how data are represented or compressed in files. It uses deep learning to detect file types accurately, and then mine Natural Redundancy for soft decoding. The second approach, called \textit{Representation-Aware}, assumes that such knowledge is known and uses it for error correction. Furthermore, both approaches combine the decoding based on NR and ECCs.

The Representation-Oblivious approach is useful for many storage systems where error correction is a low-level function. In those systems, such as hard drives or solid-state drives (SSDs), the controllers for error correction often have no access to information such as file types or compression schemes. Deep learning is a very useful tool for learning the complex patterns in data from scratch. And deep learning based classifiers are also suitable for decoding such data with Natural Redundancy. The Representation-Aware approach is useful for storage systems where error correction is a higher-level function. With knowledge on how data are represented, better error correction performance can be achieved with suitable machine learning techniques. 

This paper studies NR-based error correction for several types of data of common file types, including HTML files, JPEG files, PDF files, LaTex files and language-based texts. It presents new deep learning techniques for mining Natural Redundancy, and presents both soft-decoding and hard-decoding algorithms based on NR that can be combined with LDPC codes. It presents both experimental results and theoretical analysis for measuring the amount of Natural Redundancy mined for error correction, and the results show that NR-based decoding can substantially improve the error correction performance. (For instance, the Representation-Aware scheme can improve the decoding threshold for erasures of LDPC codes by a factor of five, when the channel's erasure rate is as high as $30\%$.) Furthermore, we also analyze the computational complexity of using Natural Redundancy for error correction versus for data compression. 

The rest of the paper is organized as follows. In Section II, we review related works. In Section III, we present the Representation-Oblivious scheme, and combine it with LDPC codes to achieve enhanced error-correction performance. In Section IV, we present a Representation-Aware scheme for language-based texts, and analyze the performance of two approaches for combining NR-based decoders with LDPC decoders: a \textit{sequential} decoding scheme and an \textit{iterative} decoding scheme. In Section V, we study the computational complexity of using NR for error correction versus for data compression. In Section VI, we present the conclusions. 

\section{Related Work}
In this section, we review related works, including joint-source channel coding (JSCC), denoising, recent results on NR-based error correction, and deep learning for information theory. 

The idea of using the leftover redundancy at a source encoder to improve the performance of ECCs has been studied within the field of joint source-channel coding (JSCC)~\cite{bauer2001variable,fresia2006combined,guivarch2000joint,hagenauer1995source,
jeanne2005joint,kim2005combined,peng2000turbo,poulliat2005analysis,pu2007ldpc}. However, few works have considered the Representation-Oblivious scheme. Furthermore, not many works have considered JSCC specifically for language-based sources.  Related to JSCC, denoising is also an interesting and well studied technique~\cite{alvarez1992image, buades2005review, chatterjee2010denoising,coifman1995translation, 
lindenbaum1994gabor, Ordentlich2008, ordentlich2003discrete, rudin1992nonlinear,
yaroslavsky1996fundamentals,DUDE2005}. A denoiser can use the statistics and features of input data to reduce its noise level for further processing. However, how to combine denoisers with the recent progress in LDPC codes and machine learning has remained under-explored.

In recent works (including results from the authors of this work), machine learning and algorithmic techniques have been used to exploit NR to correct errors in data~\cite{JiangITW2015,JiangAllerton2017,JiangITA2017,LiWaJiBrContentAsilomar2012,LuoECCcontentRecognitionWCSP2016,OnLDPC2017,DeepLearningPulakesh,YingWangISIT2017,YingWang2016}. This work studies the Representation-Oblivious scheme for the first time, and also presents new theoretical analysis for the Representation-Aware scheme.

In parallel, there have been numerous recent works on using deep learning for information theory~\cite{Ichiki,Minghai}, especially for wireless and optical communications. They mainly focus on using deep learning to model complex channels, to design codes, and to approximate or improve decoding algorithms~\cite{Aoudia,Cammerer,Dorner,HKim2,Nachmani2}. In contrast to those works, this paper focuses on using machine learning for \emph{data} with complex structures (instead of for complex channels), and on exploring error correction for such complex data. These two different directions are complementary in a communication or storage system, and can be integrated.

\section{Representation-Oblivious NR-decoding}

In this section, we study the \textit{Representation-Oblivious} scheme for Natural Redundancy (NR) based decoding. In this scheme, \textit{no prior information} on the data is needed, including how data are represented or compressed, which file type (e.g. HTML, JPEG, etc.) they belong to, or how meta-data are appended to payload bits. This scheme has the benefit of having only minimal requirements on practical storage systems such as hard drives and SSDs. Controllers of storage systems can read out blocks of data and perform error correction (aided by NR-decoding) as usual, without having to access file systems for additional information on the data. However, the task is also challenging. For example, without knowing the data compression algorithm, we cannot use its codebook to find patterns in the data. The patterns in data are highly complex, and vary greatly for different file types. (For instance, bit patterns in HTML files and JPEG files are very distinct from each other.) To address the challenges of this new error correction paradigm, we use deep learning to perform error correction in three consecutive steps: (1) detect the file type of the given block of noisy bits; (2) perform NR-based soft decoding for the block of noisy bits; (3) use the NR-based soft-decoding results to improve the performance of ECC decoding. 

\begin{figure*}
\includegraphics[height=3.9cm, width=17cm]{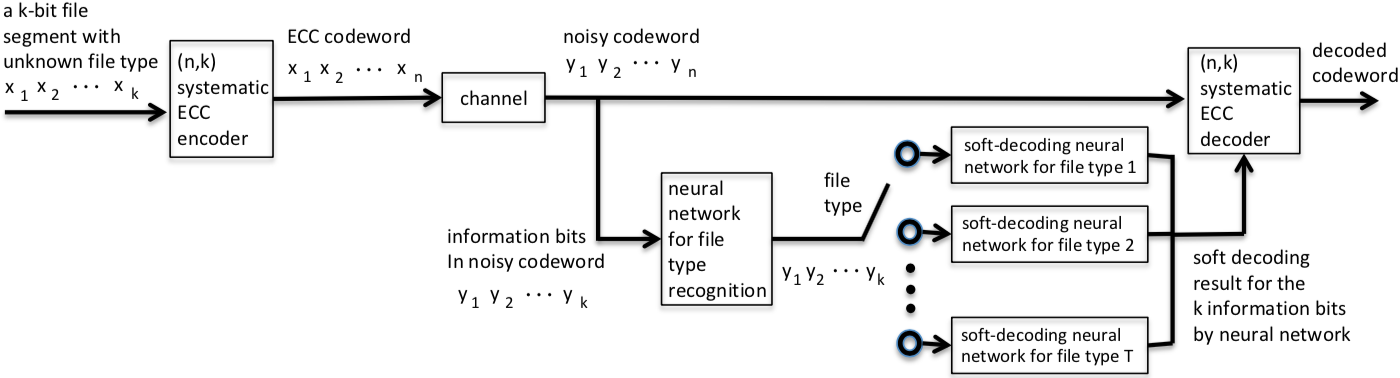}
\caption{Encoding and decoding scheme for a noisy file segment of an initially unknown file type. The $k$-bit file segment is encoded by a systematic $(n,k)$ ECC into an $n$-bit codeword. The codeword is transmitted through a channel to get a noisy codeword. Two neural networks use NR to decode the $k$ noisy information bits: the first network determines the file type of the file segment, and then a corresponding neural network for that file type performs soft decoding for the $k$ noisy information bits. The soft decoding result and the noisy codeword are both given to the ECC decoder for further error correction.}
\label{fig:codeModel}
\end{figure*}

Our coding scheme for Representation-Oblivious error correction using NR is illustrated in Fig.~\ref{fig:codeModel}. When files are stored, each file is partitioned into segments of $k$ bits, and each file segment is encoded by a systematic $(n,k)$ ECC into a codeword of $n$ bits. Then each ECC codeword passes through a noisy channel, which models the errors in a storage device. During decoding, first, a deep neural network (DNN) uses the $k$ noisy information bits to recognize the file type (e.g. HTML, LaTeX, PDF or JPEG) of the file segment. Then, a second DNN for that file type performs soft decoding on the $k$ noisy information bits based on Natural Redundancy, and outputs $k$ probabilities, where for $i=1,2,\cdots,k$, the $i$-th output is the probability for the $i$-th information bit to be 1. The $k$ probabilities are given as additional information to the ECC's decoder. The ECC decoder then performs its decoding and outputs the final result. (In our experiments, the ECC is a systematic LDPC code, and the $k$ probabilities are combined with the initial LLRs (log-likelihood ratios) for information bits to obtain their updated LLRs. The LDPC code then runs its belief-propagation (BP) decoding algorithm.) In the following, we present the detailed designs.

\subsection{File Type Recognition using Deep Learning} 
We present here a \emph{Deep Neural Network} (DNN) for file type recognition. The DNN takes a noisy file segment of $k$ bits, $(y_{1},y_{2},\cdots,y_{k})$, as input, and outputs one of $T$ file types (e.g., HTML, LaTeX, PDF or JPEG). The errors in the file segment come from a binary-symmetric channel (BSC) of bit-error rate (BER) $p$. We first introduce the architecture of the DNN and its training method. We then present the experimental results, which show that it achieves high accuracy for file type recognition.
\subsubsection{DNN Architecture and Training}
\begin{figure*}
\centering
\includegraphics[height=8cm, width=15cm]{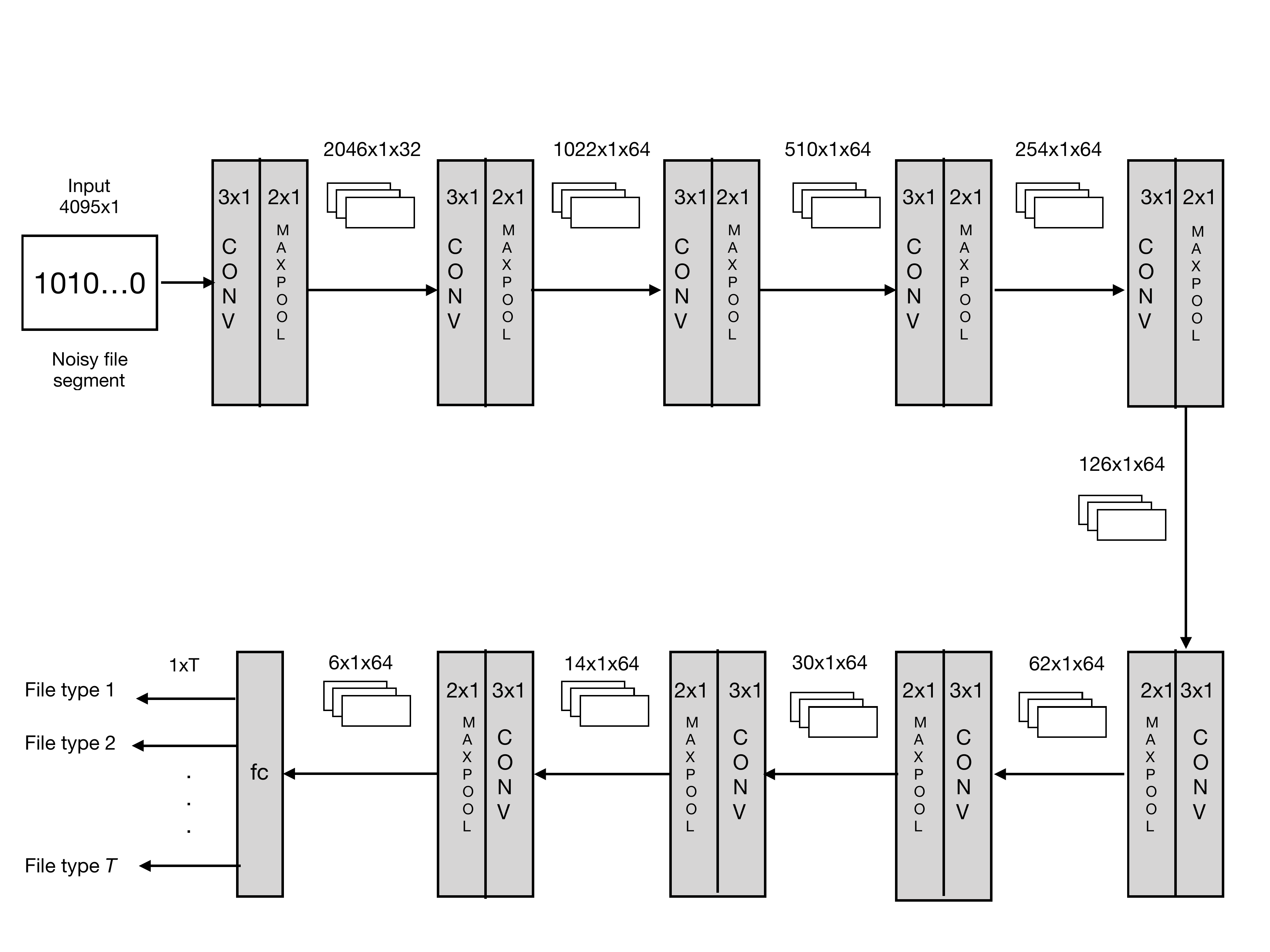}
\caption{ Architecture of the CNN (convolutional neural network) for File Type Recognition. Its input is a noisy file segment of 4095 bits, and its output corresponds to $T=4$ candidate file types (HTML, LaTex, PDF and JPEG). The CNN uses \emph{ReLU} and \emph{sigmoid} as the activation function of its convolutional layers and output layer, respectively. It uses \textit{cross entropy} as its loss function. Its optimizer is chosen to be an \textit{Ada Delta Optimizer}. }
\label{fig:FTR}
\end{figure*}
Our DNN architecture is shown in Fig.~\ref{fig:FTR}. It is a Convolutional Neural Network (CNN) that takes the $k$ bits of a noisy file segment as input. In our experiments, we let $k=4095$. (The LDPC code we use is a $(4376,4095)$ code designed by MacKay~\cite{MacKayWebpage}, which can tolerate BER of 0.2\%. Both the code length and the BER are in the typical range of parameters for storage systems.) The CNN has $T$ outputs that correspond to the $T$ possible file types, namely, the $T$ classification results. The output with the highest value leads to the selection of the corresponding file type. In our experiments, we consider four file types: HTML, LaTeX, PDF and JPEG. So $T=4$. Note that HTML and LaTeX files are both text sequences but have different file structures; PDF files contain both texts and images; and JPEG files are images. In the following, we will present DNNs and experiments using these parameters for the convenience of presentation. Note that the designs can be extended to other file-segment lengths and more file types. 

A large dataset has been used to train and test the CNN. For each of the $T=4$ file types, 24,000 noiseless file segments are used for training data, 4,000 noiseless file segments are used for validation data, and 4,800 noiseless file segments are used for test data. During training and testing, random errors of BER $p$ are added to each file segment, where each file segment uses an independently generated error pattern. 
%Since every noiseless file segment can have many noisy versions, the actual sizes of the training data, validation data and test data are significantly greater than the sizes of the noiseless data, which is effective in preventing the CNN from overfitting.

\subsubsection{Experimental Performance}

%\begin{table*} 
%\centering
%\caption{Bit error rate (BER) vs Test Accuracy for File Type Recognition (FTR). Here the ``overall test accuracy" is for all 4 types of files together. The last four columns show the test accuracy for each individual type of files. (Their average value is the overall test accuracy.)}
%\label{table1}
%\begin{tabular}{|l|l|l|l|l|l|}
%\hline 
%bit error rate (BER) & overall test accuracy & test accuracy for HTML  & test accuracy for JPEG & test accuracy for PDF & test accuracy for LaTeX  \\
%\hline 
%0.008 & 0.9969  & 0.9998 & 0.9950 &0.9935 & 0.9992 \\ 
%0.012  & 0.9966 & 0.9996 & 0.9923 & 0.9948 & 0.9996  \\
%0.016 & 0.9958 & 0.9996 & 0.9960 &0.9883 & 0.9992 \\
%0.020 & 0.9961 & 0.9995 & 0.9918 &0.9943 & 0.9991 \\
%\hline 
%\end{tabular}
%\end{table*}

%\textcolor{red}{For the experiments here for file-type recognition, can we do it for BER $p$ between 0.2\% and 0.8\%, too?}

\begin{table}
\centering
\caption{Bit error rate (BER) vs Test Accuracy for File Type Recognition (FTR). Here the ``overall test accuracy" is for all 4 types of files together. The last four columns show the test accuracy for each individual type of files. (Their average value is the overall test accuracy.)}
\label{table1}
\begin{tabular}{|l||l|l|l|l|l|} 
\hline 
\textbf{Bit Error}& \textbf{Overall} & HTML& JPEG & PDF  & LaTeX \\
\textbf{Rate} & \textbf{Test} & Test & Test & Test & Test \\
\textbf{(BER)} & \textbf{Accuracy} & Accuracy  & Accuracy & Accuracy & Accuracy \\ 
\hline 
0.2\% & 99.61\%  & 99.98\% & 99.52\% & 99.17\% & 99.77\% \\ 
0.4\% & 99.69\%  & 99.96\% & 99.60\% & 99.25\% & 99.96\% \\ 
0.6\% & 99.60\%  & 99.94\% & 99.48\% & 99.06\% & 99.90\% \\ 
0.8\% & 99.69\%  & 99.98\% & 99.50\% & 99.35\% & 99.92\% \\ 
1.2\% & 99.66\% & 99.96\% & 99.23\% &  99.48\% & 99.96\%  \\
1.6\% & 99.58\% & 99.96\% & 99.60\% &  98.83\% & 99.92\% \\
%2.0\% & 99.61\% & 99.95\% & 99.18\% &  99.43\% & 99.91\% \\
\hline 
\end{tabular}
\end{table}

The $(4376,4095)$ LDPC code used in our experiments can correct errors of BER up to 0.2\% \emph{by itself}. (That is, when it is used in the conventional way without the extra help of Natural Redundancy, it has a decoding threshold of 0.2\%.) Our goal is to use the Natural Redundancy in file segments to correct errors of substantially higher BERs. So we have selected the target BER $p$ with substantially higher values, ranging from $0.2\%$ to $1.6\%$. We then train the CNN with the given target BER $p$.

We measure the performance of the CNN by the \emph{accuracy} of file type recognition (FTR), which is defined as the fraction of file segments whose file types are recognized correctly. The test performance is shown in Table~\ref{table1}. It can be seen that file types can be recognized by the CNN with high accuracy: for all BERs, the accuracy is close to 1. 

We can also examine the accuracy for recognizing each file type, and see if there is variance in performance from file type to file type. The results are shown in the last four columns of Table ~\ref{table1}. It can be seen that overall, the accuracy is constantly high for all file types.

The CNN's performance compares favorably with existing results on FTR, which has been studied previously for applications such as disk recovery. The work~\cite{Calhoun} considered a classification method for a pair of file types using Fisher's linear discriminant and longest common subsequence methods. The accuracy ranges between $87\%$ and $99\%$ depending on which pair of file types are considered. The work~\cite{Fitzgerald} introduced an NLP (natural language processing) based method, where unigram and bigram counts of bytes and other statistics are used to generate feature representation, which is then followed by support vector machine (SVM) for classification of various file types. The classification accuracy varies from $17.4\%$ for JPEG files, $62.5 \%$ for PDF files to $94.8\%$ for HTML files. The work~\cite{Amirani} used PCA (principal component analysis) and a feed-forward auto-associative unsupervised neural network for feature extraction, and a three layer multi-layer perceptron network for classification. The classification accuracy is  $98.33\%$ for six file types while considering entire files instead of file segments. Our deep-learning based method can be seen to achieve high performance, without the need to train separate modules for feature extraction and classification. 

The CNN has \textit{robust} performance because it works well not only for the BER it is trained for, but also for other BERs in the considered range. (For example, a CNN trained for $BER=1.2\%$ also works well for other BERs in the range $[0.2\%,2.0\%]$.) For succinctness we skip the details. The robustness of the overall error correction performance for different BERs will be presented in Subsection C.

%\textcolor{red}{For the experiments here for file-type recognition, can we do it for BER $p$ between 0.2\% and 0.8\%, too?}

\subsection{Soft NR-decoding by Deep Neural Networks}
In this subsection, we study how to design DNNs that can perform soft decoding on noisy file segments. For each of the $T$ file types, we will design and train a different DNN, because different types of files have different types of Natural Redundancy. Given a file type, we will design a DNN whose input is a noisy file segment of $k$ bits 
%$X = (x_{1},x_{2},\cdots,x_{k})$
$Y = (y_{1},y_{2},\cdots,y_{k})$. As before, the errors in the noisy file segment come from a binary-symmetric channel (BSC) of bit-error rate (BER) $p$. The output of the DNN is a vector $Q=(q_{1},q_{2},\cdots,q_{k})$, where for $i=1,2,\cdots,k$, the real-valued output $q_{i} \in [0,1]$ represents the DNN's belief that for the $i$-th bit in the file segment, the probability that its correct value should be 1 is $q_{i}$. In other words, if we use 
%$B = (b_{1},b_{2},\cdots,b_{k})$ 
$X = (x_{1},x_{2},\cdots,x_{k})$ 
to denote an error-free file segment, and let it pass through a BSC of BER $p$ to obtain a noisy file segment 
%$X = (x_{1},x_{2},\cdots,x_{k})$
$Y = (y_{1},y_{2},\cdots,y_{k})$, then $q_{i}$ is the DNN's estimation for 
%$Pr\{b_{i}=1~|~X,p\}$
$Pr\{x_{i}=1~|~Y,p\}$. Note that the $k$ bits are not independent of each other because of the Natural Redundancy in them. So 
%$Pr\{b_{i}=1~|~X,p\}$ 
$Pr\{x_{i}=1~|~Y,p\}$ depends  on not only $y_i$ and $p$, but also the overall value of 
%$X$
$Y$. The goal of the DNN is to learn the Natural Redundancy in file segments, and use it to make the probability estimation $q_{i}$ be as close to 
%$Pr\{b_{i}=1~|~X,p\}$ 
the true probability $Pr\{x_{i}=1~|~Y,p\}$ 
as possible, for each $i$ and for each possible value 
%$X$ 
$Y$ of the noisy file segment. To train the DNN, our optimization objective is to minimize the loss function
$$L = \frac{1}{k} \sum\limits_{i=1}^k [x_i \log_2{q_i} + (1-x_i) \log_2(1-q_i)],$$
which measures the cross-entropy between $(x_1,x_2,\cdots,x_k)$ and $(q_1,q_2,\cdots,q_k)$, over all samples in the training dataset. 

\begin{figure*} 
\includegraphics[height=4cm, width=16cm]{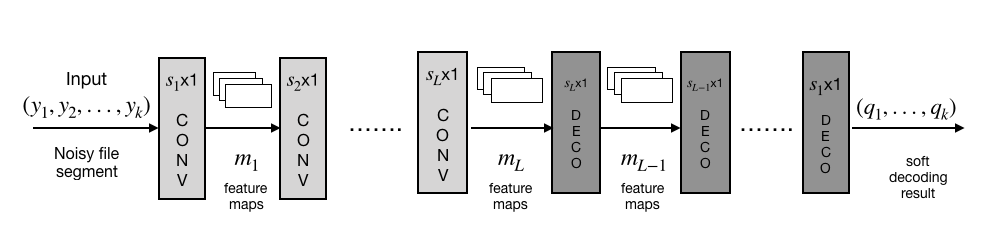}
\centering
\caption{General architecture of deep neural networks (DNNs) for NR-based soft decoding of noisy file segments. It consists of $L$ convolutional layers followed by $L$ deconvolutional layers. The activation function for the last layer is \textit{relu}, and is \textit{sigmoid }for the other layers. It uses \textit{cross-entropy} as the loss function, and uses the \textit{Adam} optimizer.}
\label{fig:ConvDeconv}
\end{figure*}

The architecture of the DNN is presented in Fig.~\ref{fig:ConvDeconv}. It is related to auto-encoders, which are good choices for various applications related to denoising~\cite{Vincent,Long}. The DNN model consists of $L$ convolutional layers followed by $L$ deconvolutional layers. (Deconvolutional layers may be seen as reverse operations of convolutional layers. Interested readers can refer to~\cite{Chollet} for more details.) The $L$ convolutional layers have one-dimensional filters of size $s_1,s_2,\cdots,s_L,$ respectively, and the number of feature maps at the output of each layer is $m_1,m_2,\cdots,m_L,$ respectively. The filter sizes and the number of feature maps for deconvolutional layers change in the reverse order.

\begin{figure*} 
\centering
\includegraphics[height=5cm, width=14cm]{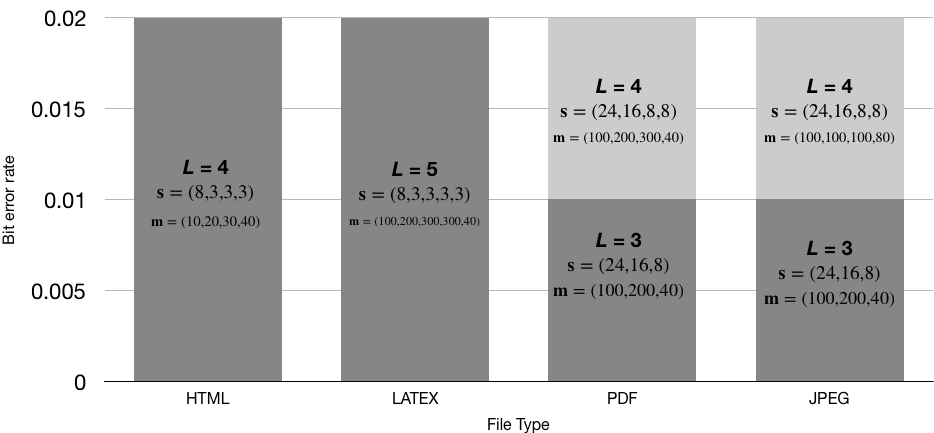}
\caption{Hyper-parameters of optimized DNN models for NR-based soft decoding, for $T=4$ file types and different bit error rates. Here $L$ is the number of convolutional/deconvolutional layers, $\mathbf{s} = (s_1,s_2,\cdots,s_L)$ represents the filter sizes, and $\mathbf{m} = (m_1,m_2,\cdots,m_L)$ represents the numbers of feature maps.}
\label{fig:Architecture}
\end{figure*} 

We optimize the hyper-parameters of DNNs (including filter sizes, number of feature maps, etc.) for each file type. Their performance is robust: an optimized DNN usually performs soft decoding well for a wide range of BERs. However, the performance can be slightly improved further if the hyper-parameters are also optimized based on BERs. Such optimization results are presented in Fig.~\ref{fig:Architecture}. (Here, for PDF and JPEG files, the hyper-parameters are optimized based on two sub-ranges of BERs.) We will present their decoding performance (when combined with ECC decoding) and robustness in the next subsection. 

\subsection{Combine Soft NR-decoding with Soft LDPC-decoding}
In this subsection, we present a scheme that combines the soft NR-decoding, which applies deep learning to noisy file segments of different file types, with soft LDPC-decoding. The experimental results confirm that the scheme substantially improves the reliability of different types of files.

We adopt a \emph{robust scheme} here: the DNNs for file-type recognition and for soft decoding have been trained with a constant BER $p_{DNN}$, but they are used for a wide range of BERs $p$ for the BSC channel. (For example, the DNNs may be trained just for $p_{DNN} = 1.2\%$, but are used for any BER $p$ from 0.2\% to 1.6\% in experiments here.) We choose this robust scheme because when DNNs are designed, the future BER in data can be highly unpredictable.

Given a noisy systematic LDPC codeword, we first use a DNN to recognize its file type based on its $k$ noisy information bits. Then a second DNN for that file type is used to do soft decoding for the $k$ noisy information bits, and output $k$ probabilities: for $i=1,2,\cdots,k$, the $i$-th output $q_i$ represents the estimated probability for the $i$-th information bit to be 1. Those $k$ probabilities can be readily turned into LLRs (log-likelihood ratios) for the information bits using the formula $$LLR_{i}^{DNN} = \log (\frac{1-q_{i}}{q_{i}})$$

For $i=1,2,\cdots,n$, let $LLR_{i}^{channel}$ be the LLR for the $i$-th codeword bit (with $1 \le i \le k$ for information bits, and $k+1 \le i \le n$ for parity-check bits) derived for the binary-symmetric channel, which is either $\log(\frac{1-p}{p})$ (if the received codeword bit is 0) or $\log(\frac{p}{1-p})$ (if the received codeword bit is 1). Then we let the \emph{initial LLR} for the $i$-th codeword bit be $$LLR_{i}^{init} = LLR_{i}^{channel} + LLR_{i}^{DNN}$$ for $1\le i\le k$, and $$LLR_{i}^{init} = LLR_{i}^{channel}$$ for $k+1\le i\le n$. 
%Here $\alpha$ is a parameter that shows the trustworthiness of the LLR derived by DNN. (In practice, we found $\alpha=1$ is a good choice, which means the soft-decoding result of the DNN is about as reliable as the LLR derived from channel.) 
We then perform belief-propagation (BP) decoding using the initial LLRs, and obtain the final result.

Note that there is a positive -- although very small -- chance that the file type will be recognized incorrectly. In that case, the incorrect soft-decoding DNN will be used, which is accounted for in the overall decoding performance for fair evaluation. We measure the performance of the error correction scheme by the percentage of codewords that are decoded correctly, which we call \emph{Decoding Success Rate}. (Let us call the scheme the \emph{NR-LDPC decoder}, since it combines decoding based on Natural Redundancy and the LDPC code.) We focus on BERs that are beyond the decoding threshold of the LDPC code, because NR becomes helpful in such cases. Note that the $(4376,4095)$ LDPC code used in our experiments has a decoding threshold of $BER=0.2\%$. In our experiments, we focus on BERs $p$ that are not only beyond the decoding threshold, but also can be significantly larger: $p \in [0.2\%,~1.6\%]$.

\begin{figure*}
\includegraphics[height=9cm, width=16cm]{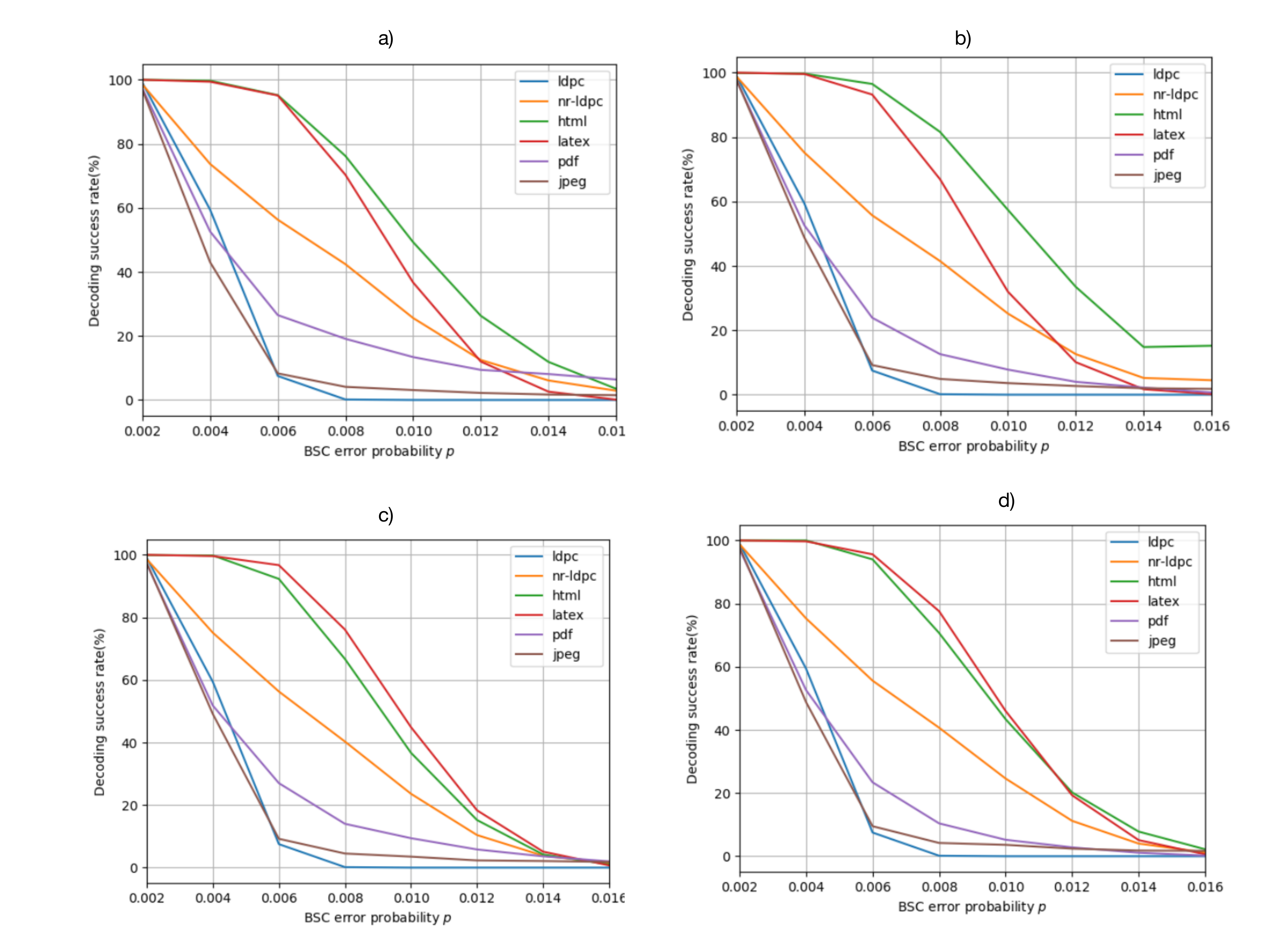}
\caption{Decoding success rate vs BER for (a) $p_{DNN}$ = $1.0\%$ , (b) $p_{DNN}$ = $1.2\%$, (c) $p_{DNN}$ = $1.4\%,$ (d) $p_{DNN}$ = $1.6\%$.} 
\label{fig:PERF}
\end{figure*}

The experimental results for $p_{DNN} = 1.0\%$ are presented in Fig.~\ref{fig:PERF} (a). Here the $x$-axis is the channel error probability $p$, and the $y$-axis is the \emph{Decoding Success Rate}. (For each $p$, 1000 file segments with independent random error patterns have been used in experiments.) The curve for ``ldpc'' is the performance of the LDPC decoder alone, and the curve for ``nr-lpdc'' is for the NR-LDPC decoder. It can be seen that the NR-LDPC decoder achieves significantly higher performance. For example, as $p=0.6\%$, the decoding success rate of the NR-LDPC decoder is approximately 4 times as high as the LDPC decoder.

The figure also shows the performance for each of the 4 file types. (The 4 curves are labelled by ``html'', ``latex'', ``pdf'', ``jpeg'', respectively. Their average value becomes the curve for ``nr--ldpc''.) It shows that the error correction performance for HTML and LaTex files are significantly better than for PDF and JPEG files. It is probably because the former two mainly consist of languages, for which the soft-decoding DNNs are better at  finding their patterns and mining their natural redundancy, while PDF is a mixture of languages and images and JPEG is image only. It is interesting to notice that even for JPEG files, when $p > 0.6\%$, the NR-LDPC decoder again performs better than the LDPC decoder, which means the DNNs can extract Natural Redundancy from images, too. Fig.~\ref{fig:PERF} (b) to Fig.~\ref{fig:PERF} (d) show the performance for $p_{DNN}$ = 1.2\%, 1.4\% and 1.6\%, respectively. The NR-LDPC decoder performs equally well in those cases, which proves the value of Natural Redundancy for decoding.
%%%%%%%%%%%%%%%%%

In summary, although no prior information is known on data representation, deep learning can recognize file types with high accuracy, and perform soft decoding effectively. When combined with ECCs, it can improve the error correction performance substantially. It is expected that with future improvements in deep learning, more natural redundancy can be mined from data to improve the reliability of storage systems even further.

\section{Representation-Aware NR-decoding}
The previous section studies \textit{Representation-Oblivious} schemes. In this section, we study schemes that are \textit{Representation-Aware}: how the source data is mapped to bits in files is known. This is also a highly useful scenario, especially when error correction is performed at a high level in computer systems, or when the controller in storage devices perform their own compression schemes. In this work, we focus on language-based data, which form an important part of big data. In particular, we focus on the English language compressed by LZW algorithms. The results can be generalized to more languages and other sequential compression algorithms, such as Huffman codes~\cite{LiWaJiBrContentAsilomar2012}
\textit{etc.}

In this section, we first present an NR-based hard-decoding algorithm for languages, and analyze its performance. We then study two important cases for combining NR-decoding with ECC decoding: the \textit{sequential} decoding scheme,  and the \textit{iterative} decoding scheme. For both cases, we study how NR-decoding improves the decoding thresholds of LDPC codes. Both the experimental results and the theoretical analysis show the ability of NR decoding to enhance the reliability of storage systems. 

\subsection{NR-decoder for Languages}
Consider English texts compressed by an LZW (Lempel-Ziv-Welch) algorithm that uses a fixed dictionary of size $2^{\ell}$. In our experiments, we use $\ell = 20$, which gives a dictionary of $2^{20}$ patterns (larger than many practical LZW codes). The dictionary has $2^{\ell}$ text strings (called patterns) of variable lengths, where every pattern is encoded as an $\ell$-bit codeword. Given a text to compress, the LZW algorithm scans it and partitions it into patterns, and maps them to codewords. For example, if we compress  \textit{``Flash memory is an $\cdots$''}, \textit{``Flash m''} gets mapped to a 20-bit codeword, \textit{``emory i''} gets mapped to another codeword and so on. The LZW code has been constructed using the Wikipedia corpus. It can compress English texts to 2.94 bits/character, which is substantially higher than the rate of 4.59 bits/character achieved by the commonly used character-level Huffman codes. The fixed dictionary of the LZW code also makes it easy to use in practice. 

In this section, we focus on bit-erasure channels. For long LZW-compressed texts with erasures, to make the NR-decoding efficient, we present a decoding algorithm based on sliding-windows of variable lengths as follows. 
 
 \subsubsection{Baseline Algorithm} 
Let $n_{min}$ and $n_{max}$ be two integers, where $n_{min} < n_{max}$ and let $\ell$ be the length of LZW codewords. We first use a sliding-window of $n_{min}\ell$ bits to scan the compressed text (where every such window contains exactly $n_{min}$ LZW codewords of size $\ell$), and obtain candidate solutions for each window based on the validity of words. (Specifically, if the bits in the window contain $t$ erasures, there are $2^t$ possible solutions, each of which can be mapped back to a text string. If all the whole words in the text string are valid words, the solution is considered a candidate solution.) We then increase the size of the window to $(n_{min}+1)\ell$, $(n_{min}+2)\ell$, $\cdots$, $n_{max}\ell$, and do decoding for each size in the following dynamic programming approach.
 
Consider a window of $k\ell$ bits that contains $k$ LZW-codewords $C_{1}$, $C_{2}$, $\cdots$, $C_{k}$. Let $S_{1} \subseteq \{0,1\}^{(k-1)\ell}$ be the set of candidate solutions for the sub-window that contains the LZW-codewords $C_{1}$, $C_{2}$, $\cdots$, $C_{k-1}$; and let $S_{2} \subseteq \{0,1\}^{(k-1)\ell}$ be the set of candidate solutions for the sub-window that contains the LZW-codewords $C_{2}$, $C_{3}$, $\cdots$, $C_{k}$. (Both $S_{1}$ and $S_{2}$ have been obtained in the previous round of decoding.) We now obtain the set of candidate solutions for the current window, which contains $C_{1}$, $C_{2}$, $\cdots$, $C_{k}$, this way. A bit sequence $(b_{1},b_{2},\cdots,b_{k\ell})$ is in $S$ only if it satisfies two conditions: (1) its first $(k-1)\ell$ bits are a solution in $S_{1}$, and its last $(k-1)\ell$ bits are a solution in $S_{2}$; (2) the decompressed text corresponding to it contains no invalid words (except on the boundaries). This way, potential solutions filtered by smaller windows will not enter solutions for larger windows, making decoding more efficient. As a final step, an erased bit is decoded this way: if \emph{any} of the windows of size $n_{max}\ell$ containing it (note that there are up to $2n_{max}-1$ such windows) can recover its value, decode it to that value; otherwise it remains as an erasure.

\subsubsection{Phrase and Word Length Filter} 
To make the above decoding algorithm more efficient, we also use phrases (such as ``information theory'', ``flash memory'') and features such as word/phrase lengths. If a solution for a window contains a valid word or phrase that is particularly long, we may remove other candidate solutions that contain only short words. That is because long words and phrases are very rare: their density among bit sequences of the same length decreases exponentially fast as the length increases~\cite{JiangITW2015}. So if they appear, the chance that they are the correct solution is high based on Bayes' rule. The thresholds for such word/phrase lengths can be set sufficiently high such that the probability of making a decoding error is sufficiently small.
\subsubsection{Co-location Filter}
\begin{figure}	
\begin{center}
\includegraphics[height = 5cm,width=12cm]{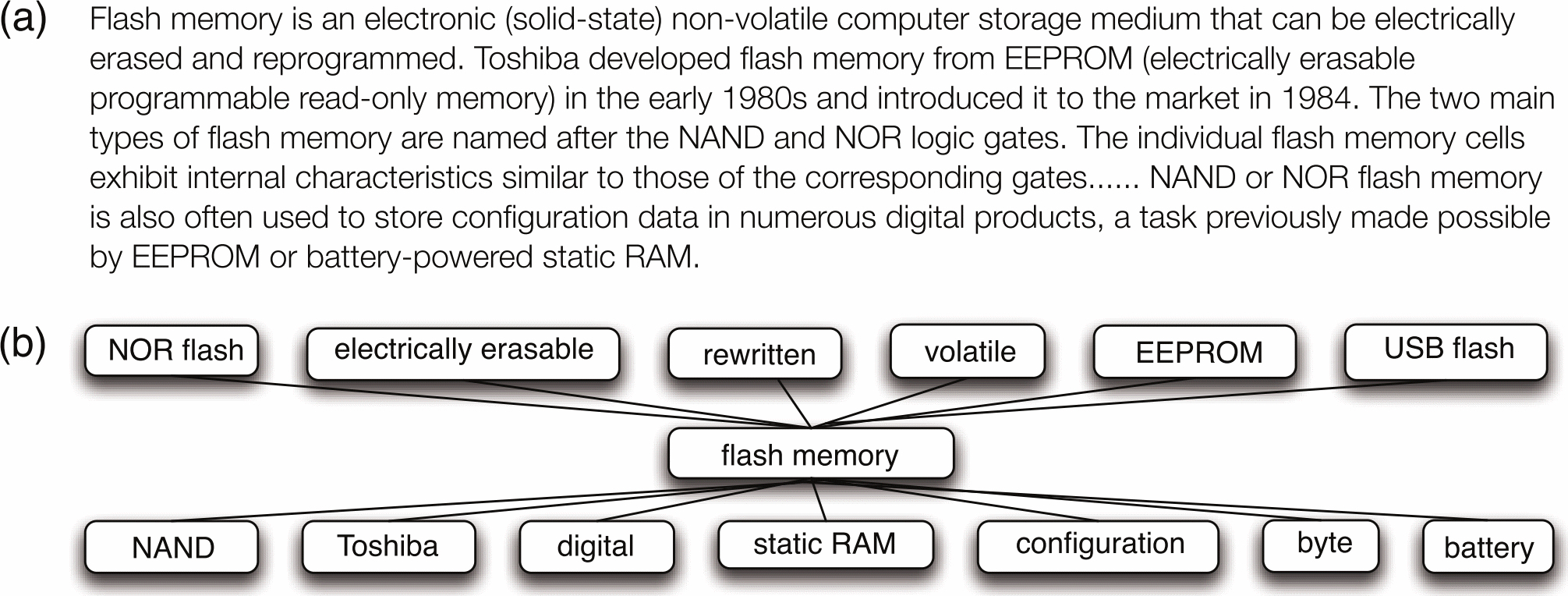} 
\end{center}
\caption{\footnotesize Co-location relationship between words and phrases. (a) A sample paragraph from Wikipedia (part of which was omitted to save space). (b) Phrases in it that have the co-location relationship with ``flash memory''.}~\label{fig:sampleRelatedTokens_v4}
\end{figure}

We also enhance the decoding performance by using the \emph{co-location} relationship. Co-location means that certain pairs of words/phrases appear unusually frequently in the same context (because they are closely associated), such as ``dog'' and ``bark'', or ``information theory'' and ``channel capacity''. If two words/phrases with the co-location relationship are detected among candidate solutions for two windows close to each other, we may keep them as candidate solutions and remove other less likely solutions. The reason for this approach is similar to that for long words/phrases. The co-location relationship can appear in multiple places in a text, and therefore help decoding in non-trivial ways. For example, for the text in Fig.~\ref{fig:sampleRelatedTokens_v4} (a), the words/phrases that have the co-location relationship with the phrase ``flash memory'' are shown in Fig.~\ref{fig:sampleRelatedTokens_v4} (b). How to find words/phrases with the co-location relationship from a corpus of training texts is a well-known technique in Natural Language Processing (NLP)~\cite{StatNLPbook}. So we skip its details here. 

We present the above decoding algorithm's performance for the binary erasure channel (BEC). The output of the NR-decoder has both erasures and errors (which will be further decoded by ECC later on). Let $\epsilon \in [0,1]$ be the raw bit-erasure rate (RBER) of BEC. After NR-decoding, for an originally erased bit, let $\delta \in [0,1]$ denote the probability that it remains as an erasure, and let $\rho \in [0,1-\delta]$ denote the probability that it is decoded to 0 or 1 incorrectly. Then the amount of noise after NR-decoding can be measured by the entropy of the noise (erasures and errors) per bit: 
$$E_{NR}(\epsilon) \triangleq \epsilon (\delta + (1-\delta) H(\frac{\rho}{1-\delta}) ),$$  where $H(p)=-p\log p-(1-p)\log(1-p)$ is the entropy function. Some typical values of $E_{NR}(\epsilon)$ are shown in Table~\ref{fig:noiseredtable}. The reduction in noise by NR-decoding is $\frac{\epsilon - E_{NR}(\epsilon)}{\epsilon}$. The table shows that noise is reduced very effectively (from 88.0\% to 91.6\%) for the LZW compressed data (without any help from ECC), for RBER from $5\%$ to $30\%$, which is a wide range for storage systems.\vspace{2mm}

%=================================================
\begin{table}[h]
\centering
\begin{tabular}[height=2.5cm,width=10cm]  {|c||c|c|c|} 
 \hline
RBER $\epsilon$ & 0.05 & 0.10 & 0.15 \\  
\hline   
\hline 
$\delta$   & $8.22 \times 10^{-2}$ & $8.67 \times 10^{-2}$ & $9.19 \times 10^{-2}$ \\
\hline
$\rho$     & $9.18 \times 10^{-5}$ & $1.83 \times 10^{-4}$ & $1.82 \times 10^{-4}$ \\
\hline
$E_{NR}(\epsilon)$ &  $4.18\times 10^{-3}$ & $8.92\times 10^{-3}$ & $1.42\times 10^{-2}$   \\
\hline
Noise     & 91.6\% & 91.1\% & 90.6\% \\
reduction &        &        &        \\
\hline
\end{tabular}
\end{table}
\begin{table}[h]
\centering
\begin{tabular}[height=2.5cm,width=10cm]  {|c||c|c|c|} 
\hline
RBER $\epsilon$ & 0.20 & 0.25 & 0.30 \\  
\hline 
\hline 
$\delta$   & $9.76 \times 10^{-2}$ & $1.05\times 10^{-1}$ & $1.12\times 10^{-1}$ \\
\hline
$\rho$     & $3.61 \times 10^{-4}$ & $4.48\times 10^{-4}$ & $7.11\times 10^{-4}$ \\
\hline
$E_{NR}(\epsilon)$ & $2.04\times 10^{-2}$ & $2.76\times 10^{-2}$ & $3.60\times 10^{-2}$   \\
\hline
Noise     & 89.8\% & 89.0\% & 88.0\% \\
reduction &        &        &        \\
\hline

\end{tabular}\vspace{2mm}
\caption{Noise reduction by NR-based language decoder for different erasure rates $\epsilon$.}
~\label{fig:noiseredtable}
\end{table}
Suppose that the LZW-codewords, seen as information bits, are protected by a systematic ECC. The NR-decoder can work collaboratively with the ECC decoder to maximize the number of correctable erasures. We now study two important cases for combining NR decoding with ECC decoding: the \textit{sequential} decoding scheme,  and the \textit{iterative} decoding scheme.

\subsection{Sequential Decoding by NR and LDPC code} 
\begin{figure} [h]
\centering
\includegraphics[height=4.5cm, width=9cm]{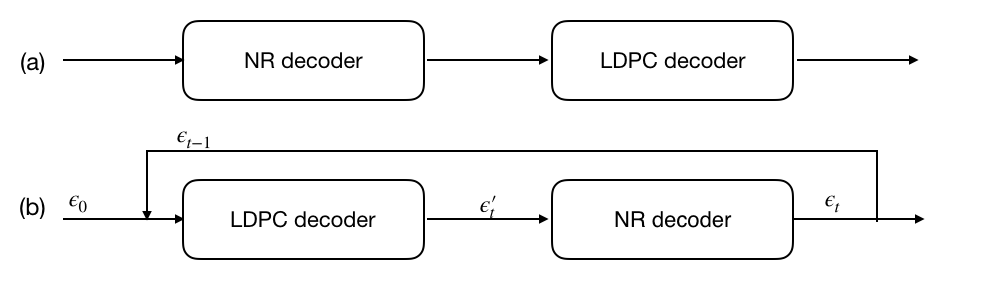}
\caption{Two schemes for combining NR-decoding with LDPC-decoding. (a) A sequential decoding scheme by NR and LDPC code. (b) An iterative decoding scheme by NR and LDPC code. }
\label{fig:schemes}
\end{figure} 

This subsection discusses the combination of NR-decoder with LDPC codes. We protect compressed text as information bits by a systematic LDPC code of rate $R$. The NR-decoder studied here generalizes the one presented in the previous subsection: it decodes the information bits by NR, and possibly the parity check bits as well using their relations with the information bits. The decoding process is a concatenation of two decoders: (1) first, the NR-decoder corrects erasures and outputs a partially corrected codeword; (2) then, the LDPC decoder takes that codeword as input (where the erasure and error probabilities result from the NR-decoding), and uses belief propagation (BP) for decoding. (See Fig.~\ref{fig:schemes} (a) for an illustration.) We present a theoretical analysis for the decoding performance, and show that the NR-decoder can substantially improve the performance of LDPC codes.

Consider a binary-erasure channel (BEC) with erasure probability $\epsilon_{0}$. Let us call the non-erased bits \emph{fixed bits}. Assume that after NR-decoding, a non-fixed bit (i.e., erasure) remains as an erasure with probability $p_{0}(\epsilon_0)\in [0,1]$, becomes an error (0 or 1) with probability $(1-p_{0}(\epsilon_0))\gamma_{0}(\epsilon_0) \in [0,1-p_{0}(\epsilon_0)]$, and is decoded correctly (as 0 or 1) with probability $(1-p_{0}(\epsilon_0))(1-\gamma_{0}(\epsilon_0))$. (In general, $p_{0}(\epsilon_0)$ and $\gamma_{0}(\epsilon_0)$ can be functions of $\epsilon_0$. Note that if the NR-decoder decodes only information bits, and an erasure in the information bits remains as an erasure with probability $p_{0}(\epsilon_0)'$, then $p_{0}(\epsilon_0)=Rp_{0}(\epsilon_0)' + (1-R)$. Also note that the LDPC decoder needs to decode all bits with both errors and erasures.)

\subsubsection{Decoding Algorithm}

We design the following iterative LDPC decoding algorithm, which generalizes both the peeling decoder for BEC and the Gallager B decoder for BSC~\cite{RyanLin}:

\begin{algr}
Generalized LDPC decoding algorithm.

1) Let $\pi \in [1,d_{v}-1]$ and $\tau \in [1,d_{v}-1]$ be two integer parameters; 

2)  In each iteration, for a variable node $v$ that is an erasure, if $\pi$ or more non-erased message bits come from $d_{v}-1$ check nodes and they all have the same value, set $v$ to that bit value; 

3)  If $v$ is not a fixed bit and not an erasure (but possibly an error) in this iteration, change $v$ to the opposite bit value if $\tau$ or more non-erased message bits come from $d_{v}-1$ check nodes and they all have that opposite value. 
(The updated value of $v$ will be sent to the remaining check node in the next iteration.) 
\end{algr}

\subsubsection{Density Evolution Analysis} 

We now analyze the density evolution for the decoding algorithm, for an infinitely long and randomly constructed LDPC code of regular degrees. 

For $t=0,1,2\cdots$, let $\alpha_{t}$ and $\beta_{t}$ be the fraction of codeword bits that are errors or erasures, respectively, after $t$ iterations of LDPC decoding. We have $\alpha_0 = \epsilon_0 (1-p_0(\epsilon_0))\gamma_0(\epsilon_0)$ and  $\beta_0 = \epsilon_0 p_0(\epsilon_0)$. Let $\kappa_0 =  \epsilon_0 (1-p_0(\epsilon_0))(1-\gamma_0(\epsilon_0))$.

\begin{thm}
	For a regular $(d_{v},d_{c})$ LDPC code with variable-node degree $d_{v}$ and check-node degree $d_{c}$, we have $$\alpha_{t+1} = \alpha_0 C_t+\kappa_0 D_t +\beta_0 \mu_t,$$ 
	where
	$$C_t = 1 - (1-A_t)^{d_v-1} + \sum_{i=0}^{\tau-1}{d_v-1 \choose i}B_t^i(1-A_t-B_t)^{d_v-i-1},$$
	$$D_t = \sum_{j=\tau}^{d_v-1} {d_v-1 \choose j}A_t^j(1-A_t-B_t)^{d_v-1-j},$$ 
	$$\mu_t = \sum_{m=\pi}^{d_v-1} {d_v-1 \choose m}A_t^m(1-A_t-B_t)^{d_v-1-m},$$ whose component variables are computed iteratively as 
	$$A_t = \frac{(1-\beta_t)^{d_c-1}-(1-\beta_t - 2 \alpha_t )^{d_c-1}}{2},$$
	$$ B_t = \frac{(1-\beta_t)^{d_c-1}+(1-\beta_t - 2 \alpha_t )^{d_c-1}}{2}.$$
	For the LDPC code, we also have
	$$\beta_{t+1} = \beta_0(1-\mu_t - \nu_t),$$ 
	where $$\nu_t =\sum_{m=\pi}^{d_v-1} {d_v-1 \choose m}B_t^m(1-A_t-B_t)^{d_v-1-m}.$$
\end{thm}

\begin{IEEEproof}
	Consider the root variable node of a computation tree. After $t$ iterations, let $A_t$ denote the probability that an incoming message to the root node from a neighboring check node is an error, and let $B_t$ denote the probability that the message is correct. Then $1-A_t-B_t$ is the probability that the message is an erasure. Let $\mu_t$ (respectively, $\nu_t$) be the probability that among the $d_{v}-1$ incoming messages from neighboring check nodes to the root node, $\pi$ or more messages are errors (respectively, correct) and the remaining messages are all erasures. In the $(t+1)$-th iteration, we can have an error in the root node in one of the following cases:
	
	\begin{enumerate}
		\item The root node was initially (namely, before decoding begins) an error (which has probability $\alpha_0$), and either of the two disjoint events happens: 1) fewer than $\tau$ check-node messages are correct and the remaining messages are all erasures, which happens with probability $\sum\limits_{i=0}^{\tau-1}{d_v-1 \choose i}B_t^i(1-A_t-B_t)^{d_v-i-1}$; 2) at least one check-node message is an error, which happens with probability $1 - (1-A_t)^{d_v-1}$. The probability that either of the two events occurs is $C_t = 1- (1-A_t)^{d_v-1}+\sum\limits_{i=0}^{\tau-1}{d_v-1 \choose i}B_t^i(1-A_t-B_t)^{d_v-i-1}$.   

		\item The root node was initially correct (which has probability $\kappa_0$), but $\tau$ or more check-node messages are errors and the rest are all erasures (which happens with probability $D_t = \sum\limits_{j=\tau}^{d_v-1} {d_v-1 \choose j} A_t^j(1-A_t-B_t)^{d_v-1-j}$).

		\item The root node was initially an erasure (which has probability $\beta_0$), and $\pi$ or more check-node messages are errors and the rest are all erasures (which happens with probability $\mu_t$).  
	\end{enumerate}
	
Therefore the error rate after $t+1$ iterations will be $\alpha_{t+1} = \alpha_0 C_t +\kappa_0 D_t + \beta_0 \mu_t$.  In the $(t+1)$-th iteration, we can correct an erasure at a root node correctly if the root node was initially an erasure, and $\pi$ or more check-node messages are correct and the rest are all erasures. This happens with probability $\beta_0 \nu_t$. The root node will remain as an erasure if it is neither corrected mistakenly nor corrected correctly. So the erasure rate after $t+1$ iterations will be $\beta_{t+1} = \beta_0(1 - \mu_t -\nu_t)$. 

Now we need to find the values of $A_t$, $B_t$, $\mu_t$ and $\nu_t$.  The incoming message from a check node to the root node is correct if out of the  $d_c-1$ non-root variable nodes connected to the check node, an even number of nodes are errors and the rest are all correct (i.e., neither errors nor erasures). That probability is $B_t =  \sum\limits_{k=0}^{\lfloor\frac{d_c-1}{2} \rfloor} {d_c -1 \choose 2k} \alpha_t^{2k} (1-\alpha_t - \beta_t)^{d_c-1-2k} = \frac{(1-\beta_t)^{d_c-1}+(1-\beta_t-2 \alpha_t )^{d_c-1}}{2} $. The incoming message from a check node to the root node is an error if out of the  $d_c-1$ non-root variable nodes connected to the check node, an odd number of nodes are errors and the rest are all correct. That probability is $ A_t =  \sum\limits_{k=1}^{\lfloor\frac{d_c}{2} \rfloor} {d_c -1 \choose 2k-1} \alpha_t^{2k-1} (1-\alpha_t - \beta_t)^{d_c-2k} = \frac{(1-\beta_t)^{d_c-1}-(1-\beta_t-2 \alpha_t )^{d_c-1}}{2} $. 
The probability that $\pi$ or more neighboring check-node messages are errors and the rest are all erasures can be simplified as $\mu_t =  \sum_{m=\pi}^{d_v-1} {d_v-1 \choose m}A_t^m(1-A_t-B_t)^{d_v-1-m}$. The probability that $\pi$ or more neighboring check-node messages are correct and the rest are all erasures can be simplified as $\nu_t =\sum_{m=\pi}^{d_v-1} {d_v-1 \choose m} B_t^m(1-A_t-B_t)^{d_v-1-m} $. This completes the proof.
\end{IEEEproof}
\subsubsection{Erasure Threshold}      
Define \emph{erasure threshold} $\epsilon^{*}$ as the maximum erasure probability (for $\epsilon_{0}$) for which the LDPC code can decode successfully (which means the error/erasure probabilities $\alpha_{t}$ and $\beta_{t}$ both approach 0 as $t\to \infty$). Let us show how the NR decoder can substantially improve $\epsilon^{*}$. Consider a regular LDPC code with $d_{v}=5$ and $d_{c}=100$, which has rate 0.95 (a typical code rate for storage systems). Without NR-decoding, the erasure threshold is $\tilde{\epsilon}^{*} = 0.036$. Now let $\pi=1$ and $\tau=4$. For LZW-compressed texts, when $\epsilon_{0}=0.2$, the NR-decoder in the previous subsection gives $p_{0}=0.143$ and $\gamma_{0}=0.0003$, for which the LDPC decoder has $\lim_{t\to\infty}\alpha_{t}=0$ and $\lim_{t\to\infty}\beta_{t}=0$. (The same happens for $\epsilon_{0}< 0.2$.) So with NR-decoding, $\epsilon^{*} \ge 0.2$, which means the improvement in erasure threshold is more than $455.6\%$.

\subsection{Iterative Decoding by NR and LDPC code}

In this subsection, we study the decoding performance when we use \emph{iterative decoding} between the LDPC decoder and NR-decoder, as shown in Fig.~\ref{fig:schemes} (b). (In last subsection's study, the NR-decoder is followed by the LDPC decoder, without iterations between them.) As before, we focus on languages and systematic LDPC codes, and present a theoretical model for compressed languages as follows.

Let $\mathbb{T}=(b_{0},b_{1},b_{2},\cdots)$ be a compressed text. Partition $\mathbb{T}$ into segments $S_{0},S_{1},S_{2}\cdots$, where each segment $S_{i}=(b_{il},b_{il+1},\cdots,b_{il+l-1})$ has $l$ bits. Consider erasures in the compressed text. Let $\theta \in [0,1]$, $l_{\theta}\triangleq \lfloor l\theta\rfloor$ and $p\in [0,1]$ be parameters. We assume that when a segment $S_i$ has at most $l_{\theta}$ erasures, the NR-decoder can decode it by checking the validity of up to $2^{l_{\theta}}$ candidate solutions (based on the validity of their corresponding words/phrases, grammar, etc.), and either determines (independently) the correct solution with probability $p$ or makes no decision with probability $1-p$. (Note that an NR-decoder does not have to check the $2^{l_{\theta}}$ candidate solutions one by one. For example, the NR-decoder introduced earlier can remove many invalid solutions early on without exhaustive search.) And this NR-decoding operation can be performed \emph{only once} for each segment (because if the correct solution cannot be determined by such an NR-based operation the first time, there is no guarantee that such operations in the future will find the correct solution).  

The parameter $l_{\theta}$ here is used to bound the computational complexity and erasure-correction capability of the NR-decoder in the worst case, and $p$ models the probability of making an error-free decision. This is a simplification of the practical NR-decoder shown in the previous subsection that makes very high-confidence -- although not totally error-free -- decisions. The model is suitable for compression algorithms such as LZW coding with a fixed dictionary, Huffman coding, etc., where each segment can be decompressed to a piece of text. The greater $l$ is, the better the model is.
 \subsubsection {Iteration with LDPC Decoder}
The compressed text $\mathbb{T}$ is protected as information bits by a systematic LDPC code. The LDPC code uses the peeling decoder for BEC (where $d_{c}-1$ incoming messages of known values at a check node determine the value of the outgoing message on the remaining edge) to correct erasures. See the decoding model in Fig.~\ref{fig:schemes} (b). In each iteration, the LDPC decoder runs \emph{one iteration} of BP decoding, then the NR-decoder tries to correct those $l$-information-bit segments that contain at most $l_{\theta}$ erasures (if those segments were never decoded by the NR-decoder in any of the previous iterations). Let $\epsilon_{0}<1$ be the BEC's erasure rate. Let $\epsilon_{t}'$ and $\epsilon_{t}$ be the LDPC codeword's erasure rate after the $t$-th iteration of the LDPC decoder and the NR-decoder, respectively. Next, we analyze the density evolution for regular $(d_{v},d_{c})$ LDPC codes of rate $R$ = $1-\frac{d_v}{d_c}$. 

Note that since the NR-decoder decodes only information bits, for the LDPC decoder, the information bits and parity-check bits will have different erasure rates during decoding. Furthermore, information bits consist of $l$-bit segments, while parity-check bits do not. For such an $l$-bit segment, if the NR-decoder can decode it successfully when it has no more than $l_{\theta}$ erasures, let us call the segment \emph{lucky}; otherwise, call it \emph{unlucky}. Lucky and unlucky segments will have different erasure rates during decoding, too. 

Every $l$-information-bit segment is \emph{lucky} with probability $p$, and \emph{unlucky} with probability $1-p$. A lucky segment is guaranteed to be decoded successfully by the NR-decoder once the number of erasures in it becomes less than or equal to $l_{\theta}$; and an unlucky segment can be considered as \emph{never} to be decoded by the NR-decoder (because such decoding will not succeed). Since whether a segment is lucky or not is independent of the party-check constraints and the LDPC-decoder, for analysis we can consider it as an inherent property of the segment (which exists even before the decoding begins). 

\subsubsection{Density Evolution Analysis}

Define $q_0 = 1$, $q_t \triangleq \frac{\epsilon_t}{\epsilon_t'}$ and $d_{t}\triangleq \frac{\epsilon_{t}'}{\epsilon_{t-1}}$ for $t \ge 1$. Note that decoding will end after $t$ iterations if one of these conditions occurs: (1) $\epsilon_{t}' =0$, because all erasures are corrected by the $t$-th iteration; (2) $d_{t}=1$, because the LDPC decoder corrects no erasure in the $t$-th iteration, and nor will the NR-decoder since the input codeword is identical to its previous output. We now study density evolution before those boundary cases occur.

For $t=1,2,3\cdots$ and $k=0,1,\cdots,l$, let $f_k(t)$ denote the probability that a lucky segment contains $k$ erasures after $t$ iterations of decoding by the NR-decoder. 

\begin{lem}
\[
 f_k(1) = \begin{cases}
   \sum \limits_{i=0}^{l_\theta} {l\choose i} (\epsilon_1')^i(1-\epsilon_1')^{l-i} \mbox{~~~~~~if~} k=0 \\
0 \mbox{~~~~~~~~~~~~~~~~~~~~~~~~~~~~~~~if~} 1 \le k \le l_{\theta} \\     
{l\choose k} (\epsilon_1')^k(1-\epsilon_1')^{l-k} \mbox{~~~~~~~~~if~} l_{\theta}+1 \le k \le l 
 \end{cases}
\]  
\end{lem}
\begin{IEEEproof}
Consider the LDPC-decoding and the NR-decoding in the first iteration. Since the initial erasure rate is $\epsilon_0$, the erasure rate after LDPC decoding will now be 
	$\epsilon_1'= q_0 \epsilon_0 (1-(1-\epsilon_0)^{d_c-1})^{d_v-1}$ where $q_0 = 1$ by definition. The probability that an $l$-information-bit segment contains exactly $i$ erasures is given by ${l}\choose{i}$ $(\epsilon_1')^i(1-\epsilon_1')^{l-i}$, which is independent of whether the segment is lucky or unlucky. Thus the probability that a \emph{lucky} segment contains up to $l_\theta$ erasures is given by $\sum_{i=0}^{l_\theta}$ ${l}\choose{i}$ $(\epsilon_1')^i(1-\epsilon_1')^{l-i}$. All such segments are decoded by the NR-decoder successfully, while the remaining segments are not. That leads to the conclusion.
\end{IEEEproof}

\begin{lem}
The erasure rate after the first iteration of NR-decoding is 
$$\epsilon_{1} = \epsilon_0 d_1((1-R)+R(1-p))+( \sum_{k=l_{\theta}+1}^{l} \frac{k}{l} f_k(1))Rp$$
\end{lem}

\begin{figure} 
	\begin{center}
	\includegraphics[width=8cm]{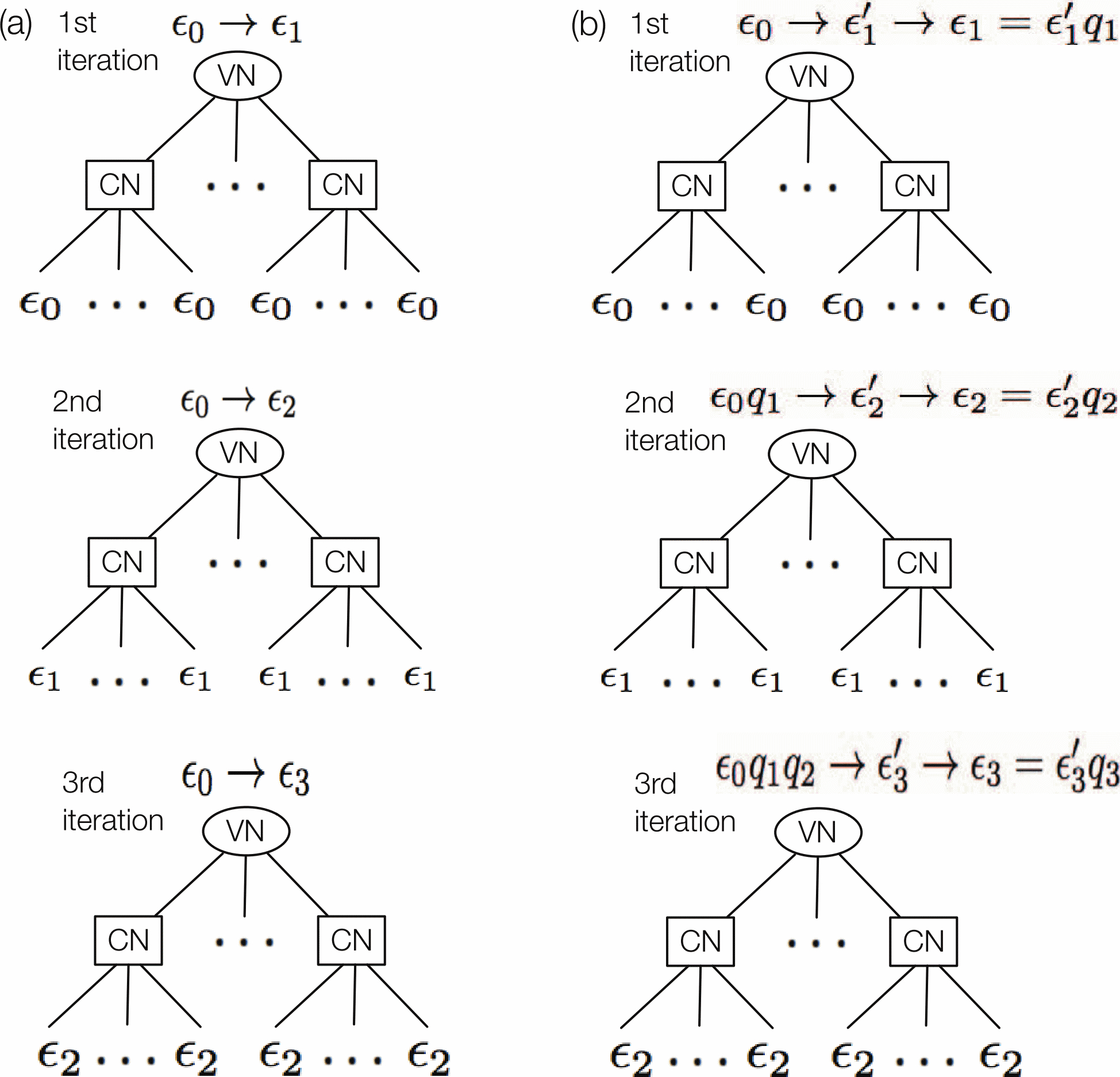} 
	\end{center}
	\caption{\footnotesize Comparison of the computation tree for density evolution analysis. (a) First three iterations of classic BP decoding (alone) for BEC. (b) First three iterations of BP-decoding and NR decoding.}~\label{fig:computationTrees}
\end{figure}

\begin{IEEEproof}
After NR-decoding, the erasure rate of a lucky segment with $k$ erasures is $\frac{k}{l}$, and the erasure rate for unlucky segments and parity-check bits is still $\epsilon_1'$. We have $d_1 = \epsilon_1'/\epsilon_0$. Hence the overall erasure rate after the 1st iteration of NR-decoding is $\epsilon_{1} = \epsilon_0 d_1((1-R)+R(1-p))+( \sum_{k=l_{\theta}+1}^{l} \frac{k}{l} f_k(1))Rp$. (See Fig.~\ref{fig:computationTrees} (b) for an illustration of the computation tree for density evolution. For comparison, we show the tree for classic BP decoding for BEC in Fig.~\ref{fig:computationTrees} (a).)
\end{IEEEproof}

\begin{lem}\label{lem:2ndLDPC}
The erasure rate after the second iteration of LDPC-decoding is  $$\epsilon_2'= q_0 q_1 \epsilon_0 (1-(1-\epsilon_1)^{d_c-1})^{d_v-1}.$$
\end{lem}

\begin{IEEEproof}
We have $q_1 = \frac{\epsilon_1}{\epsilon'_1}$. Since the NR-decoding of the 1st iteration reduces the \emph{overall} erasure probability by a factor of $q_{1}$ (from $\epsilon_{1}'$ to $\epsilon_{1}$), and the root variable node of a computation tree is chosen uniformly at random from the infinitely long and randomly constructed LDPC code, the root node in the tree for the 2nd iteration of LDPC decoding now has the erasure probability $q_1 \epsilon_0$. (See Fig.~\ref{fig:computationTrees} (b).) Hence the equation for the LDPC-decoder for the 2nd iteration will be given by $\epsilon_2'= q_0 q_1 \epsilon_0 (1-(1-\epsilon_1)^{d_c-1})^{d_v-1}$. Note that LDPC decoding is independent of NR-decoding because the parity-check constraints are independent of the bits being lucky-segment bits, unlucky-segment bits or parity-check bits. And note that $d_{2} = \frac{\epsilon_2'}{\epsilon_1}$ is the probability that an erasure \emph{remains as an erasure} after the LDPC decoding. If $d_2 = 1$, no change was made by the LDPC-decoder;  if $d_2 = 0$, all erasures have been corrected. In both cases, the decoding will end. 
\end{IEEEproof}
\begin{lem}
For $t \ge 2$, 
\[
f_k(t) = \begin{cases}
f_k(t-1)+ \sum \limits_{i=l_\theta+1}^{l} \sum \limits_{j=0}^{l_\theta} f_i(t-1){i \choose j} (d_{t})^{j}(1-d_{t})^{i-j}, \mbox{~~~~~~~~~if~} k=0 \\
\\
0, \mbox{~~~~~~~~~~~~~~~~~~~~~~~~~~~~~~~~~~~~~~~~~~~~~~~~~~~~~~~~~~~~~~~~~if~} 1 \le k \le l_{\theta} \\ 
\sum \limits_{i=k}^{l} f_i(t-1) {i \choose k}(d_{t})^{k}(1-d_{t})^{i-k}, \mbox{~~~~~~~~~~~~~~~~~~~~~~~~~~~~~~if~} l_{\theta}+1 \le k \le l  
 \end{cases}
\]      
\end{lem}
\begin{IEEEproof}
Now consider the second iteration of NR-decoding. We only consider the case when $0 < d_{2} < 1$. A lucky segment has zero errors after the second iteration if an only if either one of the two cases happen : 1) the segment already has zero errors after the first iteration, or 2) the segment had $l_\theta+1$ or more errors after the first iteration and it has at most $l_\theta$ erasures after second iteration of the LDPC-decoding. Thus if $k = 0$, 
$$f_k(2) = f_k(1)+\sum \limits_{i=l_\theta+1}^{l} \sum \limits_{j=0}^{l_\theta} f_i(1){i \choose j} (d_2)^{j}(1-d_2)^{i-j}$$
A lucky segment cannot have $k \le l_\theta$ erasures (with $k\ge 1$) after the second iteration of NR-decoding (because if so, it would have corrected those erasures). So we have $f_k(2) = 0$ for that case. 
Finally, a lucky segment has $l_\theta +1 \le k \le l$ erasures if and only if it had $k$  or more erasures after the first iteration of NR-decoding and it has $k$ erasures after the second iteration of LDPC-decoding. Thus
$$ f_k(2) =\sum \limits_{i=k}^{l} f_i(1) {i \choose k}(d_2)^{k}(1-d_2)^{i-k} \textrm{  if $l_{\theta}+1 \le k \le l$}$$

The remaining cases can be analyzed similarly. That leads to the conclusion.
\end{IEEEproof}   

We now present the analytical formulas for the density evolution of the iterative LDPC-NR decoding scheme. Its proof follows the previous lemmas.

\begin{thm} 
	For $t \ge 1$,
\[\epsilon_{t} = ((1-R)+R(1-p))\epsilon_0(\prod\limits_{i=1}^{t} d_t)+Rp\sum\limits_{k=l_\theta+1}^{l} \frac{k}{l}f_k(t),\]
\[\epsilon'_{t} = (\prod_{m=0}^{t-1} q_m) \epsilon_0(1-(1-\epsilon_{t-1})^{d_c-1})^{d_v-1}.\]                                                                                                       
\end{thm}
\begin{IEEEproof}
%	Let $f_k(t)$ denote the probability that a lucky segment contains $k$ erasures after $t$ iterations of decoding by the NR-decoder.
The decoding performance for the 2nd iteration of the LDPC-decoding has been analyzed in Lemma~\ref{lem:2ndLDPC}. The erasure rate in unlucky-segment bits and parity-check bits was decreased from $\epsilon_1'$ to $\epsilon_1' d_2$ = $\epsilon_0 d_1 d_2$ by the LDPC-decoding. Now the NR-decoder corrects those lucky segments that had more than $l_{\theta}$ erasures before the LDPC-decoding but now has at most $l_{\theta}$ erasures after the LDPC-decoding. So 
$\epsilon_{2} = \epsilon_0 d_1 d_2 ((1-R)+R(1-p))+( \sum \limits_{k=l_{\theta}+1}^{l} \frac{k}{l} f_k(2))Rp$.

The analysis for the following iterations is similar to the 2nd iteration. In general, since in the $i$-th iteration the NR-decoder reduces the overall erasure rate by a factor of $q_{i}$, the root variable node in the computation tree for the $t$-th iteration of LDPC decoding has the erasure probability $(\prod_{i=0}^{t-1}q_{i})\epsilon_{0}$. That leads to the conclusion. 
\end{IEEEproof}
   
\subsubsection{Performance}
We now numerically show that the iterative NR-LDPC decoder can improve the decoding threshold for erasures significantly. Note that the analysis in this subsection is based on the assumption that the NR-decoder corrects erasures but does not create errors. However, all our existing NR-decoders still create errors with small probabilities (such as $10^{-4}$ in Table II) which, although small, are still non-zero due to the complexity of languages. Extending the NR-decoders here to correct both erasures and errors is beyond the scope of this paper. Therefore, the following analysis is based on the same assumption as above, and the parameters of the NR-decoder are chosen reasonably based on existing experimental evidence: let each segment have $l =120$ bits (which corresponds to 6 LZW codewords of 20 bits each); and let $l_\theta=30$. (Note that in the experiments of the previous two subsections, sliding windows of the same size and more erasures have been considered.) Let the LDPC code be a regular code with $d_v= 5$ and $d_c= 100$. 

\begin{figure} [h]
	\begin{center}
	\includegraphics[height = 7cm, width=15cm]{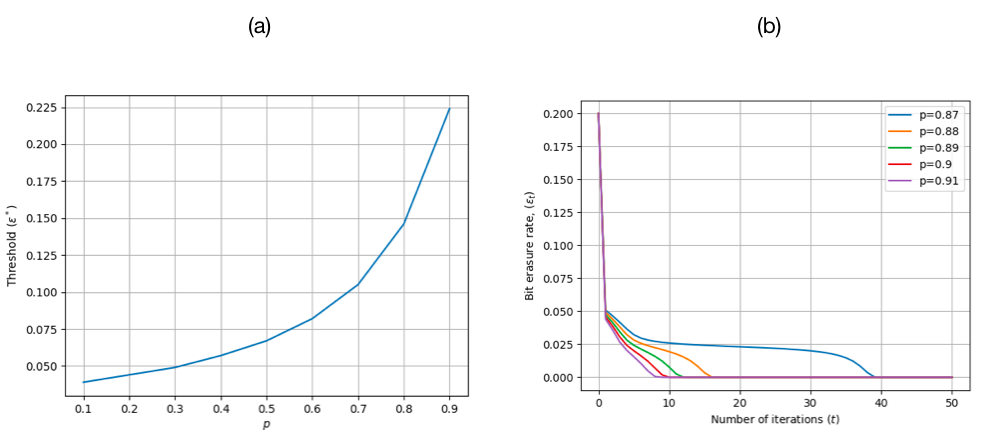} 
	\end{center}
	\caption{\footnotesize Performance of the iterative NR-LDPC decoding. (a) Here parameter $p$ is the probability that the NR-decoder corrects erasures in a segment when it has at most $l_\theta$ erasures, and parameter $\epsilon^*$ is the decoding threshold for erasures of the iterative NR-LDPC decoder. The figure shows that $\epsilon^*$ increases rapidly as $p$ increases, and it can substantially outperform the decoding threshold of the LDPC code alone (which is 0.036). (b) Here $t$ is the number of iterations of the iterative NR-LDPC decoding process, and $\epsilon_t$ is the overall bit erasure rate of the LDPC codeword after the $t$-th iteration. The bit erasure rate of the BEC channel is set to be $\epsilon_0 = 0.2$. The figure shows that the higher $p$ is, the more quickly decoding ends.}
	~\label{fig:iterations}
\end{figure}

Recall that $p$ is the probability that an NR-decoder can correct the erasures in a segment successfully when the segment has at most $l_{\theta}$ erasures. Based on the previous analysis, given the value of $p$, we can obtain the corresponding decoding threshold for erasures for the iterative NR-LDPC decoder. The results are shown in Fig.~\ref{fig:iterations} (a). It can be seen that as $p$ increases, the decoding threshold $\epsilon^*$ increases quickly. Note that without the NR-decoder, the decoding threshold of the LDPC code alone for erasures is $\tilde{\epsilon}^* = 0.036$. In Fig.~\ref{fig:iterations} (a), the decoding threshold increases from 0.039 to 0.224, all of which are higher than $\tilde{\epsilon}^*$. Based on Table II, it is reasonable to consider $p=0.9$. In this case, the decoding threshold is $\epsilon^* = 0.224$, which represents a $522.22\%$ increase from $\tilde{\epsilon}^*$. 

We also study how quickly decoding converges in the iterative decoding scheme. The results are shown in Fig.~\ref{fig:iterations} (b). Here, the BER of the BEC channel is $\epsilon_0 = 0.2$ (which is above the decoding threshold of the LDPC code alone). It can be seen that decoding converges faster as $p$ increases. In particular, when $p = 0.9$, it takes only about 7 iterations for decoding to converge. 

In summary, with knowledge on how data are represented by bits, effective NR-based decoding schemes can be designed. Both sequential and iterative schemes are presented for combining NR-decoders with LDPC codes, and their performance is rigorously analyzed. The results show that the inclusion of NR-decoding can improve LDPC decoding substantially, and iterative decoding between the two decoders can further improve performance effectively.

\section{Computational-Complexity Tradeoff for NR-based Coding}
In the Introduction section, we have mentioned that the Natural Redundancy in data can be used for both compression and error correction.  How to use it suitably depends on many factors, such as available coding techniques, hardware design, etc. In this chapter, we discuss one such tradeoff of \textit{central importance}: the computational complexity of using NR for compression or error correction. Real NR is hard to model precisely, so we explore this topic from a theoretical point of view, and consider NR in general forms. We show that certain types of redundancy are computationally efficient for compression, while others are so for error correction. Note that there exist works on analyzing the hardness of certain types of source coding schemes~\cite{LinVectorImage1992,LinStorerCohn1992,ruhlHartensteinFractal} and channel coding schemes~\cite{BruckNaor1990, DumerMicciSudan2003, FeigeMicci2004, VardySTOC1997, VardyTransIT1997}. In contrast, here we focus on the tradeoff between the two, and the analysis is NR-oriented.    

Let $B= (b_{1},b_{2},\cdots,b_{n})\in \{0,1\}^{n}$ be an $n$-bit message with NR. Define $\mathcal{V}~:~\{0,1\}^{n}\to \{0,1\}$ as a \emph{validity function}: $B$ is a valid message if and only if $\mathcal{V}(B)=1$. The set of all valid messages of $n$ bits is $\mathcal{M} \triangleq \{B\in \{0,1\}^{n}~|~\mathcal{V}(B)=1\}$. For simplicity, for both source and channel coding, assume that the valid messages in $\mathcal{M}$ are equally likely.
        
First, consider source coding. Let $k~=~\lceil \log_{2}|\mathcal{M}| \rceil$. Define an \emph{optimal lossless compression scheme} to be an injective function $C_{opt}~:~\mathcal{M} \to \{0,1\}^{k}$ that compresses any valid message $B\in \mathcal{M}$ to a distinct $k$-bit vector $C_{opt}(B)$. Define the \emph{Data Compression Problem} as follows:
Given a validity function $\mathcal{V}$,
%$\mathcal{V}~:~\{0,1\}^{n}\to \{0,1\}$, 
find an injective function $C_{opt}~:~\mathcal{M} \to \{0,1\}^{k}$.

Next, consider channel coding. Assume that a valid message $X=(x_{1},x_{2},\cdots,x_{n})\in \mathcal{M}$ is transmitted through a binary-symmetric channel (BSC), and is received as a noisy message $Y=(y_{1},y_{2},\cdots,y_{n})\in \{0,1\}^{n}$. Maximum likelihood (ML) decoding requires us to find a message $Z=(z_{1},z_{2},\cdots,z_{n})\in \mathcal{M}$ that minimizes the Hamming distance $d_{H}(Y,Z)$. Define the \emph{Error Correction Problem} as follows: 
	Given a validity function $\mathcal{V}$ 
 %   $\mathcal{V}~:~\{0,1\}^{n}\to \{0,1\}$ 
	and a message $Y \in \{0,1\}^{n}$, find a valid message $Z \in \mathcal{M}$ that minimizes the Hamming distance $d_{H}(Y,Z)$.
	
	Let $\mathcal{F}$ be the set of all functions from the domain $\{0,1\}^{n}$ to the codomain $\{0,1\}$. (We have $|\mathcal{F}|=2^{2^{n}}$.) 
	%Note that for the Data Compression Problem and the Error Correction Problem (which we shall denote by $\mathcal{P}_{dc}^{\mathcal{F}}$ and $\mathcal{P}_{ec}^{\mathcal{F}}$, respectively),  $\mathcal{V}$ can be any function in $\mathcal{F}$. 
	The function $\mathcal{V}$ represents NR in data. In practice, different types of data have different \emph{types} of NR. Let us define the latter concept formally. For any subset $\mathcal{T} \subseteq \mathcal{F}$, let $\mathcal{T}$ be called a \emph{type} of validity functions (which represents a type of NR). When $\mathcal{V}$ can only be a function in $\mathcal{T}$ (instead of $\mathcal{F}$), we denote the Data Compression Problem and the Error Correction Problem by $\mathcal{P}_{dc}^{\mathcal{T}}$ and $\mathcal{P}_{ec}^{\mathcal{T}}$, respectively. 
	%(So $\mathcal{P}_{dc}^{\mathcal{T}}$ and $\mathcal{P}_{ec}^{\mathcal{T}}$ are sub-problems of $\mathcal{P}_{dc}^{\mathcal{F}}$ and $\mathcal{P}_{ec}^{\mathcal{F}}$, respectively.)
The hardness of the problems $\mathcal{P}_{dc}^{\mathcal{T}}$ and $\mathcal{P}_{ec}^{\mathcal{T}}$ depends on $\mathcal{T}$. Let $S_{dc=NP,ec=P}$ denote the set of types $\mathcal{T}$ (where each type is a subset of $\mathcal{F}$) for which the data compression problem $\mathcal{P}_{dc}^{\mathcal{T}}$ is NP-hard while the error correction problem $\mathcal{P}_{ec}^{\mathcal{T}}$ is polynomial-time solvable. Similarly, let $S_{dc=P,ec=NP}$ (or $S_{dc=P,ec=P}$, $S_{dc=NP,ec=NP}$, respectively) denote the set of types $\mathcal{T}$ for which $\mathcal{P}_{dc}^{\mathcal{T}}$ is polynomial-time solvable while $\mathcal{P}_{ec}^{\mathcal{T}}$ is NP-hard (or $\mathcal{P}_{dc}^{\mathcal{T}}$ and $\mathcal{P}_{ec}^{\mathcal{T}}$ are both polynomial-time solvable, or both NP-hard, respectively). The following theorem shows that there exist validity-function types for each of those four possible cases.

\begin{thm}
	The four sets $S_{dc=NP,ec=P}$, $S_{dc=P,ec=NP}$, $S_{dc=P,ec=P}$ and $S_{dc=NP,ec=NP}$ are all non-empty.
\end{thm}
\begin{IEEEproof}
We first prove that $S_{dc=NP,ec=P} \ne \emptyset$, namely, there exists a validity-function type $\mathcal{T}_{NP,P}\subseteq \mathcal{F}$ that makes the data compression problem $\mathcal{P}_{dc}^{\mathcal{T}_{NP,P}}$ be NP-hard while making the error correction problem $\mathcal{P}_{ec}^{\mathcal{T}_{NP,P}}$ be polynomial-time solvable.
	
We define a validity function $\mathcal{V}_{NP,P}~:~\{0,1\}^{n}\to\{0,1\}$ as follows, which takes $n$ binary variables $b_{1},b_{2},\cdots,b_{n}$ as its input. Let $f_{3SAT}(b_{1},b_{2},\cdots,b_{n-1})$ be a 3-SAT Boolean formula, which is in the Conjunctive Normal Form (CNF) where each clause contains 3 variables (such as $(b_{1} \vee \bar{b}_{2} \vee \bar{b}_{3})\wedge (b_{2}\vee b_{4} \vee b_{6})\wedge (\bar{b}_{2}\vee b_{3}\vee b_{5})\wedge \cdots$, where $\vee$ is the OR operation, $\wedge$ is the AND operation, and $\bar{x}$ is the NOT of the Boolean variable $x$). Define a function $f_{even}(b_{1},b_{2},\cdots,b_{n})$ as follows: $f_{even}(b_{1},b_{2},\cdots,b_{n})$ equals 1 if $\sum_{i=1}^{n}b_{i}$ is even, and equals 0 otherwise. Similarly, define a function $f_{odd}(b_{1},b_{2},\cdots,b_{n})$ as follows: $f_{odd}(b_{1},b_{2},\cdots,b_{n})$ equals 1 if $\sum_{i=1}^{n}b_{i}$ is odd, and equals 0 otherwise. Finally, define the validity function $\mathcal{V}_{NP,P}(b_{1},b_{2},\cdots,b_{n})$ as $\mathcal{V}_{NP,P}(b_{1},b_{2},\cdots,b_{n}) \triangleq (f_{3SAT}(b_{1},b_{2},\cdots,b_{n-1}) \wedge f_{even}(b_{1},b_{2},\cdots,b_{n})) \vee f_{odd}(b_{1},b_{2},\cdots,b_{n})$. (The validity-function type $\mathcal{T}_{NP,P}$ is the set of all specific forms for the function $\mathcal{V}_{NP,P}$. Note that the same holds for the types $\mathcal{T}_{P,NP}$, $\mathcal{T}_{P,P}$ and $\mathcal{T}_{NP,NP}$ to be discussed later.)
                                                    
Given the validity function $\mathcal{V}_{NP,P}$, we can see that the set of valid messages $\mathcal{M}$ has cardinality $|\mathcal{M}| = |\{B\in \{0,1\}^{n}~|~\mathcal{V}_{NP,P}(B)=1 \}| \ge    
|\{B\in \{0,1\}^{n}~|~f_{odd}(B)=1 \}| = 2^{n-1} $, because all the messages whose bits have odd parity must be valid. So whether $|\mathcal{M}| > 2^{n-1}$ or not (which means whether $k~=~\lceil \log_{2}|\mathcal{M}| \rceil > n-1$ or not) depends on whether the 3-SAT formula $f_{3SAT}(b_{1},b_{2},\cdots,b_{n-1})$ has a satisfying solution: if there is a satisfying solution to $b_{1},b_{2},\cdots,b_{n-1}$ that makes $f_{3SAT}(b_{1},b_{2},\cdots,b_{n-1})$ be 1, then we can let $b_{n} = \oplus_{i=1}^{n-1}b_{i}\in \{0,1\}$ (where $\oplus$ is the exclusive-OR operation), which gives us an $n$-bit message of even parity that is valid (because here $f_{3SAT}(b_{1},b_{2},\cdots,b_{n-1}) \wedge f_{even}(b_{1},b_{2},\cdots,b_{n}) = 1 \wedge 1 = 1 = \mathcal{V}_{NP,P}(b_{1},b_{2},\cdots,b_{n})$), so $k > n-1$ (which means $k=n$); otherwise, there is no valid message of even parity, so $k=n-1$. So determining whether $k=n$ or $n-1$ is equivalent to solving the 3-SAT Problem $f_{3SAT}(b_{1},b_{2},\cdots,b_{n-1})$, which is a known NP-complete problem. To solve the data compression problem $\mathcal{P}_{dc}^{\mathcal{T}_{NP,P}}$, it is necessary to know the value of $k$. So the data compression problem $\mathcal{P}_{dc}^{\mathcal{T}_{NP,P}}$ is NP-hard.
                         
Consider the error-correction problem $\mathcal{P}_{ec}^{\mathcal{T}_{NP,P}}$ given the same validity function $\mathcal{V}_{NP,P}$. Given an input noisy message $Y=\{0,1\}^{n}$, we compute $\mathcal{V}_{NP,P}(Y)$. If $\mathcal{V}_{NP,P}(Y)=1$, then $Y$ is valid and we let $Z=Y$ be the decoded message; otherwise, since all messages of odd parity are valid, we just need to flip any bit of $Y$ to get a valid message $Z$ of odd parity. In both cases, we have minimized the Hamming distance $d_{H}(Y,Z)$ (which is either 0 or 1). So the error correction problem $\mathcal{P}_{ec}^{\mathcal{T}_{NP,P}}$ is polynomial-time solvable. So $S_{dc=NP,ec=P} \ne \emptyset$.                                                          

Next, we prove that $S_{dc=P,ec=NP} \ne \emptyset$. Let $H$ be an $r\times n$ binary matrix of rank $r < n$. Define the validity function $\mathcal{V}_{P,NP}~:~\{0,1\}^{n}\to\{0,1\}$ as follows: $\mathcal{V}_{P,NP}(b_{1},\cdots,b_{n}) = 1$ if and only if $H\cdot (b_{1},\cdots,b_{n})^{T} \equiv \mathbf{0} \mod 2$. (That is, the valid messages form a linear code.) Then the data compression problem $\mathcal{P}_{dc}^{\mathcal{T}_{P,NP}}$ becomes polynomial-time solvable: we can view $H$ as the parity-check matrix of an ECC, find its corresponding generator matrix and use it to compress any valid $n$-bit message into a distinct vector of $k=n-r$ bits (e.g., through Gaussian elimination). Its details are well known in coding theory, so we skip them here. The error correction problem $\mathcal{P}_{ec}^{\mathcal{T}_{P,NP}}$ is the same as the ML decoding problem of linear codes, which is known to be NP-hard~\cite{BruckNaor1990}. So $S_{dc=P,ec=NP} \ne \emptyset$.                                                  

To prove that $S_{dc=P,ec=P} \ne \emptyset$, we can let the validity function $\mathcal{V}_{P,P}(b_{1},\cdots,b_{n}) = 1$ for all inputs. In this case, all messages are valid, so both data compression and error correction become trivial problems. So $S_{dc=P,ec=P} \ne \emptyset$.   

Now we prove that $S_{dc=NP,ec=NP} \ne \emptyset$. Let $f_{3SAT}(b_{1},\cdots,b_{n})$ be a 3-SAT Boolean formula as defined before (except that here it takes $n$ bits, instead of $n-1$ bits, as input). Let function $f_{0}(b_{1},\cdots,b_{n})$ be defined this way: it equals 1 if $b_{1}=b_{2}=\cdots=b_{n}=0$, and 0 otherwise. Let the validity function be $\mathcal{V}_{NP,NP}(b_{1},\cdots,b_{n}) = f_{3SAT}(b_{1},\cdots,b_{n}) \vee f_{0}(b_{1},\cdots,b_{n})$.

For the data compression problem $\mathcal{P}_{dc}^{\mathcal{T}_{NP,NP}}$, $k > 0$ (namely, $|\mathcal{M}| > 1$) if and only if $f_{3SAT}(b_{1},\cdots,b_{n})$ has a satisfying solution whose bits are not all zeros, which is NP-complete to determine. So $\mathcal{P}_{dc}^{\mathcal{T}_{NP,NP}}$ is NP-hard.

For the error correction problem $\mathcal{P}_{ec}^{\mathcal{T}_{NP,NP}}$, let the input noisy message $Y\in \{0,1\}^{n}$ be $Y=(1,1,\cdots,1)$. Then $\min_{Z\in \mathcal{M}}d_{H}(Y,Z) < n$ if and only if $f_{3SAT}(b_{1},\cdots,b_{n})$ has a satisfying solution whose bits are not all zeros, which is NP-complete to determine. So $\mathcal{P}_{ec}^{\mathcal{T}_{NP,NP}}$ is NP-hard. So $S_{dc=NP,ec=NP} \ne \emptyset$.
\end{IEEEproof}                                                                             
The above result shows a wide range of possibilities for the computational-complexity trade-off between source and channel coding. In practice, it is worthwhile to study the properties of Natural Redundancy (e.g., whether the redundancy is mainly local or global, which differs for different types of data), and choose appropriate coding schemes based on computational complexity along with other important factors.

\section{Conclusion}
This paper explores the use of Natural Redundancy in data for error correction. It presents new NR-decoders, which are based on deep learning and machine learning, and combines them with ECC decoding. For storage systems accommodating big data, the vast amount of Natural Redundancy offers the opportunity to improve data reliability significantly. Two important paradigms are studied in the paper. In the Representation-Oblivious paradigm, no information on data representation is needed \textit{a priori}. In the Representation-Aware paradigm, both sequential and iterative decoding schemes are analyzed. The experimental and analytical results verify that machine learning can mine Natural Redundancy effectively from complex data, and improve error correction substantially. The usage of Natural Redundancy for error correction also adds minimal overhead for big storage systems, since it does not require the modification of existing data. 

\end{document}